\documentclass[a4paper,11pt]{article}
\usepackage{pos}
\usepackage{hyperref}
\usepackage{lineno}
\title{$\rm{\Lambda^{+}_{c}}$ production cross section in pp and p--Pb collisions down to $p_{\rm T}$ = 0 at $\sqrt{s_{\rm NN}}$ = 5.02 TeV measured with ALICE}
\ShortTitle{$\rm{\Lambda^{+}_{c}}$ cross section at $p_{\rm T}$ = 0 with ALICE}

\manuallySeparateAuthors
\author*[a,b]{Annalena Sophie Kalteyer}
\author{ for the ALICE Collaboration}

\affiliation[a]{GSI Helmholtzzentrum für Schwerionenforschung GmbH,\\
  Planckstraße 1, Darmstadt, Germany}

\affiliation[b]{Physikalisches Institut, Ruprecht-Karls-Universität Heidelberg,\\
Im Neuenheimer Feld 226, Heidelberg, Germany}

\emailAdd{annalena.sophie.kalteyer@cern.ch}
\abstract{
The open heavy-flavour hadron measurements in proton--proton and proton--lead collisions give insight into the charm production and hadronization mechanisms. In this contribution, the latest measurements of the production cross section of prompt $\Lambda^+_{\rm c}$ and its charge conjugate performed with the ALICE detector at midrapidity in pp, and the new measurement of $\Lambda_\mathrm{c}^+ \to \mathrm{p} \mathrm{K}_\mathrm{S}^0$ performed down to $p_{\rm T}=0$ in p--Pb collisions at $\sqrt{s_{\rm NN}}$ = 5.02 TeV are presented. We also present the first ALICE measurement of the baryon-to-meson ratio $\Lambda^+_{\rm c}/{\rm D^0}$ and the $\Lambda^+_{\rm c}$ nuclear modification factor $R_{\rm pPb}$ down to $p_{\rm T}$ = 0 in p--Pb collisions. The $\Lambda^+_{\rm c}/{\rm D^0}$ ratio at midrapidity at the LHC is significantly higher than the one in ${\rm e^+e^-}$ collisions, suggesting that the fragmentation of charm is not universal across different collision systems. The results are compared with theoretical calculations.}

\FullConference{%
 
 The Ninth Annual Conference on Large Hadron Collider Physics - LHCP2021\\
 7-12 June 2021\\
 Online
}

\begin{document}
\maketitle

\section{Introduction}

The production cross sections of open heavy-flavour hadrons are typically described within the factorisation approach as the convolution of the parton distribution functions of the incoming protons, the perturbative QCD partonic cross section, and the fragmentation functions. The fragmentation functions are parametrised from measurements in ${\rm e^+e^-}$ collisions, assuming a universality across different collision systems. Thus, measurements of charm-baryon production are crucial to study the charm quark hadronization in pp and p--Pb collisions and its difference with respect to ${\rm e^+e^-}$ collisions. Especially the baryon-to-meson ratio $\Lambda^{+}_\mathrm{c}/\mathrm{D}^0$ is sensitive to the charm hadronization mechanism. In the recent ALICE measurements an enhanced $\Lambda^{+}_\mathrm{c}$ baryon production was observed with respect to $\mathrm{e}^+\mathrm{e}^-$ collisions \cite{LcpppPb}. This could hint to further colour reconnection string topologies due to a larger number of multi-parton interactions or to the formation of a hot deconfined medium in small collision systems which would allow hadronization via coalescence. Furthermore, measurements of charm-baryon production in p--Pb collisions provide important information about Cold Nuclear Matter (CNM) effects quantified in the nuclear modification factor $R_{\mathrm{pPb}}$:
\begin{equation}
    R_{\mathrm{pPb}}=\frac{\text{d}\sigma_{\mathrm{pPb}}/\text{d}p_\mathrm{T}}{A\cdot\text{d}\sigma_{\mathrm{pp}}/\text{d}p_\mathrm{T}},
\end{equation}
where $\text{d}\sigma_{\mathrm{pPb}(\mathrm{pp})}/\text{d}p_\mathrm{T}$ are the $p_\mathrm{T}$-differential cross sections in p--Pb and pp collisions at a given centre-of-mass energy, and $A=208$ is the lead mass number. The charm hadron production measurement could also help us to understand how the possible presence of collective effects could modify the production of heavy-flavour hadrons in different colliding systems.\\
In this contribution recent ALICE results on the $\Lambda^{+}_\mathrm{c}$ baryon production in pp and p--Pb collisions will be shown \cite{LcpppPb}, as well as a new measurement of the  $\Lambda_\mathrm{c}^+ \to \mathrm{p} \mathrm{K}_\mathrm{S}^0$ production cross section down to $p_\mathrm{T}=0$ in p--Pb collisions at $\sqrt{s_{\rm NN}}$ = 5.02 TeV. The extension of the measured $p_\mathrm{T}$ range in p--Pb collisions is possible, since for the first time in ALICE the $\Lambda^{+}_\mathrm{c}$ candidates are reconstructed employing the KFParticle package \cite{KFParticle}, together with machine learning techniques. The package is based on the Kalman filter method, and it is especially suitable for short-lived particles like the $\Lambda^{+}_\mathrm{c}$, which decay before reaching the innermost detectors of the ALICE apparatus. Furthermore, particle identification and topological selections are optimized using the machine learning algorithm XGBoost \cite{XGBoost}. The extracted signal is corrected for the detector acceptance and reconstruction efficiency, and contributions from the decay of beauty hadrons are subtracted. Due to the extended $p_\mathrm{T}$ range for $\Lambda^{+}_\mathrm{c}$ baryons in p--Pb collisions a measurement of $\Lambda^{+}_\mathrm{c}/\mathrm{D}^0$ and $ R_{\mathrm{pPb}}$ down to $p_\mathrm{T}=0$ is presented.

\section{Production cross section and baryon-to-meson ratio}
 \begin{figure}[ht!]
 \centering
\includegraphics[width=0.4\textwidth]{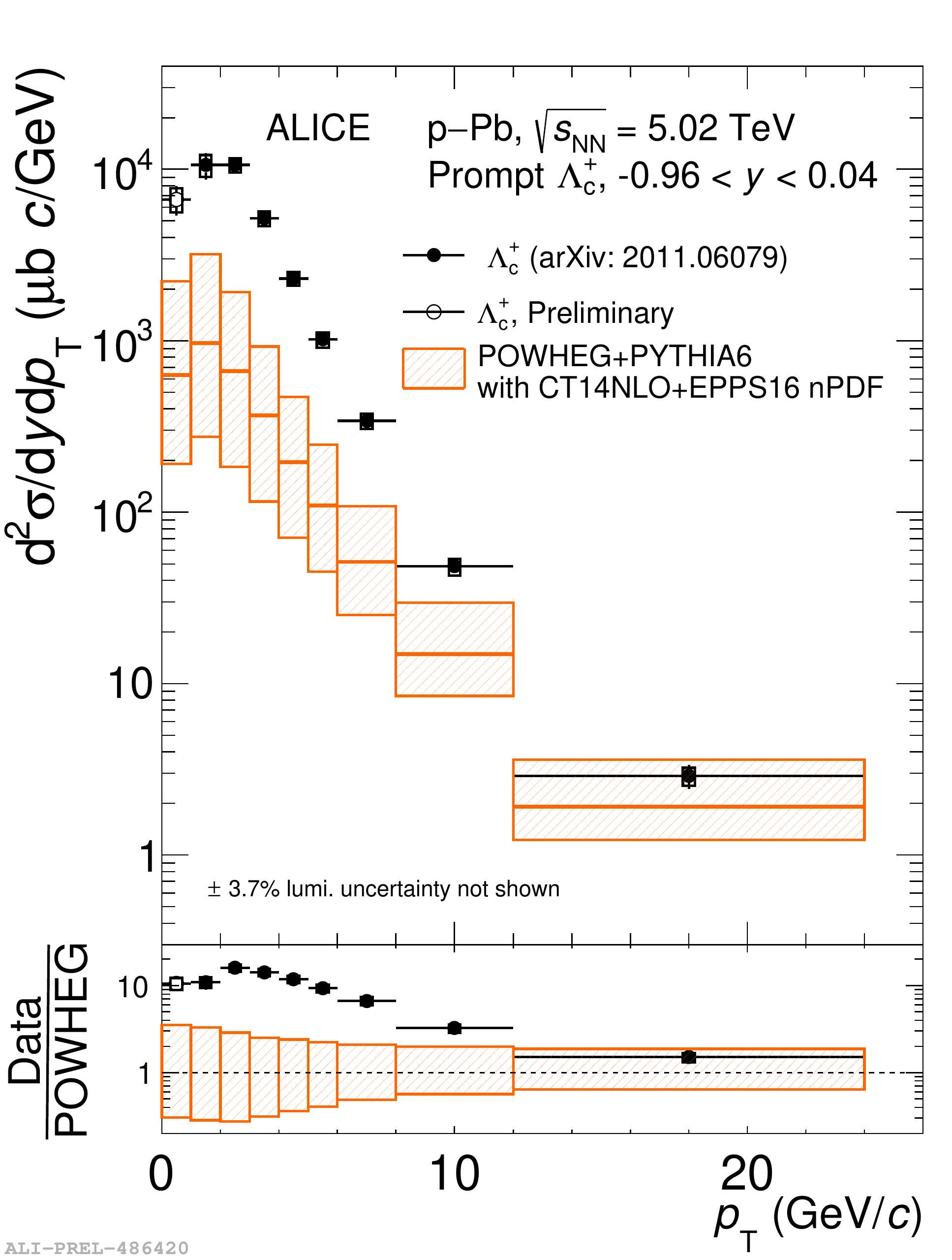}
\includegraphics[width=0.5\textwidth]{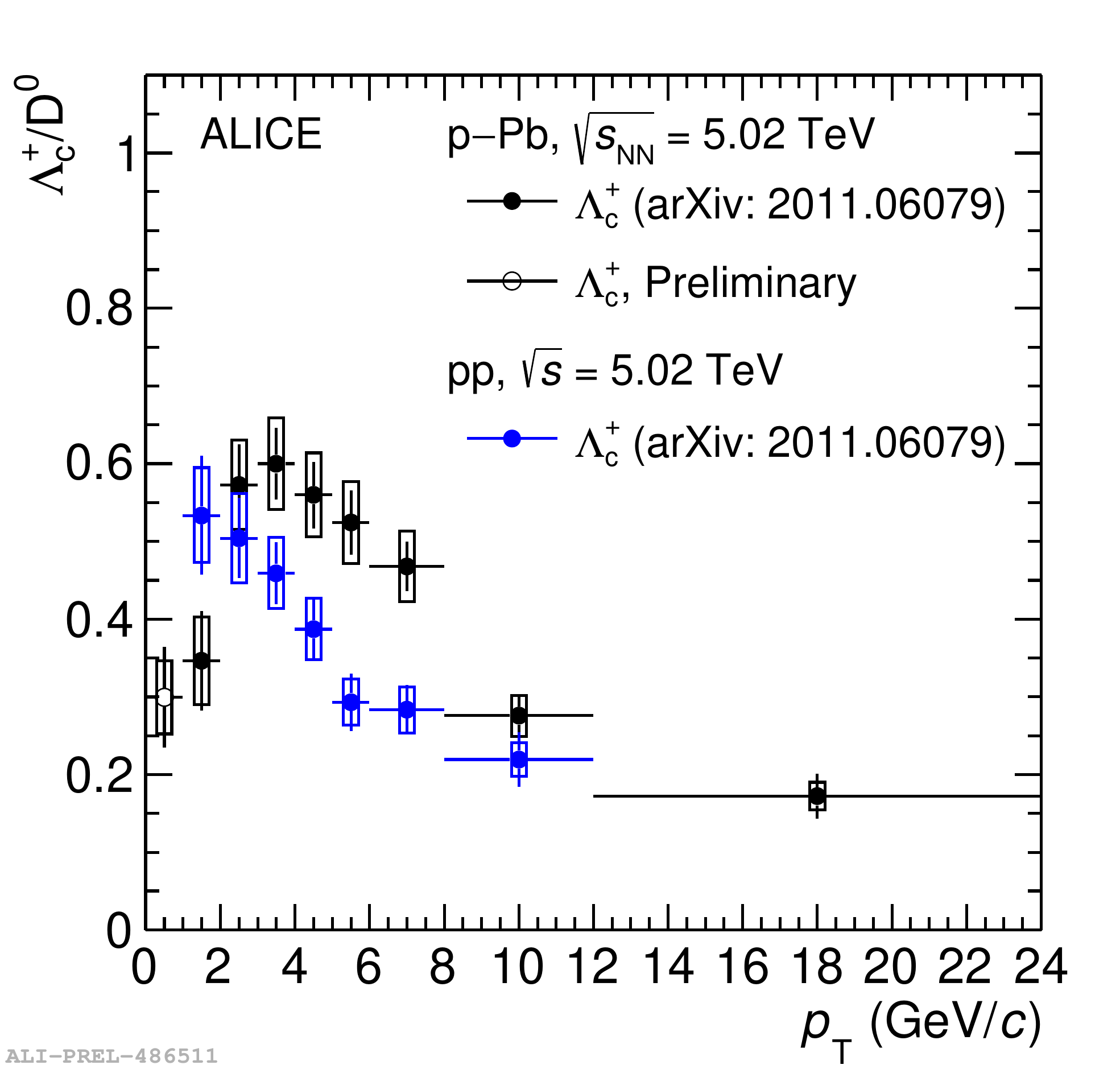}
\caption{Left: Prompt $\Lambda^{+}_\mathrm{c}$ cross section in p--Pb collisions at $\sqrt{s_{\rm NN}}$ = 5.02 TeV \cite{LcpppPb} in comparison with \textsc{POWHEG} event generator \cite{POWHEG} plus PYTHIA~$6$ model calculation \cite{PYTHIA6}. Right: $\Lambda_\mathrm{c}^+/\mathrm{D}^0$ ratio in pp and p--Pb collisions.}
\label{fig:Crosssection}
\end{figure}
 The measurement of the $\Lambda^{+}_\mathrm{c}$ production cross section is performed with the ALICE detector at a nucleon-nucleon centre-of-mass energy of $\sqrt{s_{\mathrm{NN}}} = 5.02\,$TeV. The published $p_\mathrm{T}$-differential cross section in p--Pb collisions in the range $1<p_\mathrm{T}<24$\,GeV/$c$ \cite{LcpppPb} is extended with the new measurement down to $p_\mathrm{T}=0$. In Fig.~\ref{fig:Crosssection} the result is compared to a model calculation with the \textsc{POWHEG} event generator \cite{POWHEG} and PYTHIA~$6$ \cite{PYTHIA6} for the parton shower generation using the parton distribution functions CT$14$NLO \cite{CT14NLO}. The nuclear modifications of the PDFs in p--Pb collisions are modelled with the EPPS\,$16$ nPDF parametrisation \cite{EPPS}. This model is based on the factorisation approach, using fragmentation functions tuned on data from $\mathrm{e}^+\mathrm{e}^-$ collisions. As shown in the left panel of Fig.~\ref{fig:Crosssection} the central prediction of the model significantly underestimates the data for $p_\mathrm{T} < 8$\,GeV/$c$.\\
The relative abundances of the charm meson species are sensitive to the fragmentation functions, and thus to the hadronization process. The baryon-to-meson $\Lambda_\mathrm{c}^+/\mathrm{D}^0$ ratios are shown in the right panel of Fig.~\ref{fig:Crosssection}, where the measurements in pp and p-Pb collisions at $\sqrt{s_{\rm NN}}$ = 5.02 TeV are compared. An enhancement is observed in both collision systems compared to $\mathrm{e}^+\mathrm{e}^-$ collisions ($\Lambda^{+}_\mathrm{c}/\mathrm{D}^0 = 0.113 \pm 0.013 \pm 0.006$ \cite{epluseminus}), suggesting that the charm fragmentation is not a universal process across different collision systems. Furthermore, a shift of the $p_\mathrm{T}$ distribution in p--Pb collisions with respect to pp collisions is observed. This could be an indication of radial flow in p--Pb collisions or of a multiplicity dependence of the charm quark hadronization. A hardening of the $p_\mathrm{T}$ spectra is predicted by models in the presence of a deconfined medium and also observed in the strange sector \cite{LcD0Strange}. Thus, this behaviour for heavy-flavour hadrons could also be interpreted as a hint of collective behaviour in p--Pb collisions. 

\section{Nuclear modification factor}
\begin{figure}[ht!]
 \centering
\includegraphics[width=0.9\textwidth]{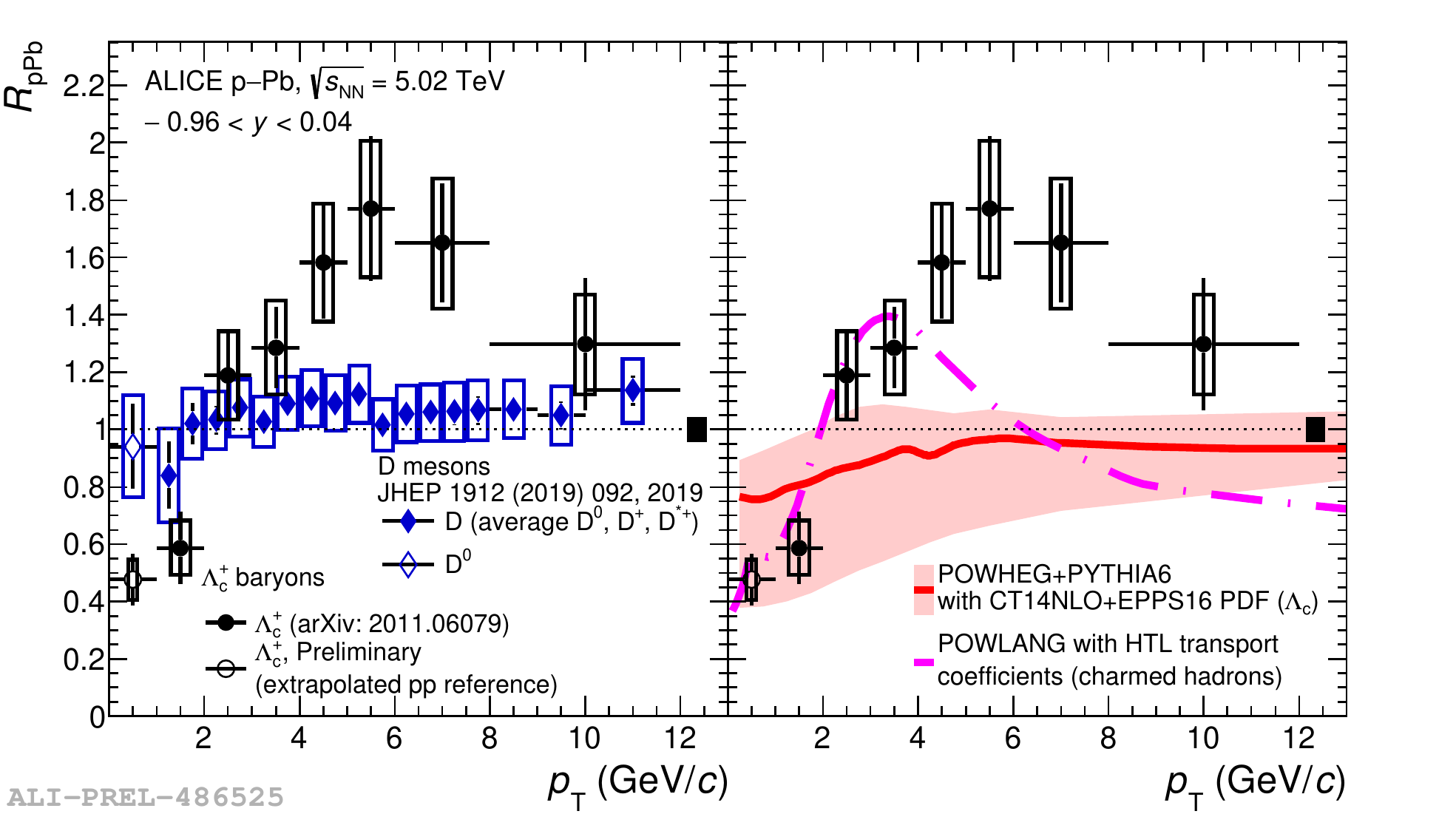}
\caption{Nuclear modification factor $R_{\mathrm{pPb}}$ for prompt $\Lambda_\mathrm{c}^+$ in p--Pb collisions at $\sqrt{s_{\mathrm{NN}}} = 5.02\,$TeV in the transverse momentum range $0$ $< p_{\mathrm{T}} <$ $12$ GeV/$c$ \cite{LcpppPb}. Left: Comparison to non-strange D mesons \cite{DMesons}. Right: Comparison of the measurement to \textsc{POWHEG} + PYTHIA~$6$ \cite{POWHEG}, \cite{PYTHIA6}, \textsc{POWLANG} model calculations \cite{POWLANG}.}
\label{fig:RpPb}
\end{figure}
The nuclear modification factor $R_{\mathrm{pPb}}$ provides a comparison of the production cross section in pp and p--Pb collisions scaled by the lead mass number. The $\Lambda_\mathrm{c}^+$ cross section in pp collisions was obtained by the ALICE Collaboration in the range $1<p_\mathrm{T}<12$\,GeV/$c$ \cite{LcpppPb} and extrapolated to lower and higher transverse momenta using a PYTHIA~$8$ model calculation \cite{PYTHIA8}. In the left panel of Fig.~\ref{fig:RpPb} the nuclear modification factor for prompt $\Lambda_\mathrm{c}^+$ in p--Pb collisions at $\sqrt{s_{\mathrm{NN}}} = 5.02\,$TeV is compared to non-strange D-mesons \cite{DMesons}. While the $R_{\mathrm{pPb}}$ is consistent with unity for the mesons, a suppression is observed at low $p_\mathrm{T}$ for $\Lambda_\mathrm{c}^+$ and an enhancement at intermediate $p_\mathrm{T}$. The right panel of Fig.~\ref{fig:RpPb} shows a comparison of the nuclear modification factor with different models. The above mentioned model, using the \textsc{POWHEG} event generator and PYTHIA~$6$, assumes hadronization via fragmentation and takes into account initial-state effects for the $\mathrm{c}\bar{\mathrm{c}}$ production, that occur in the collision of a proton with a heavy nucleus. While the model agrees with the measurement within the uncertainties at low $p_\mathrm{T}$, a tension can be seen at intermediate and high $p_\mathrm{T}$. Additionally, the shape of the distribution is not captured. The \textsc{POWLANG} transport model \cite{POWLANG} takes into account initial-state effects as well, but it assumes the formation of a hot deconfined medium in p--Pb collisions implementing hadronization via fragmentation and recombination of a charm quark with light quarks in the medium. Since the model assumes the formation of a medium the heavy quark will loose energy in the medium, which leads to a suppression of the nuclear modification factor at low $p_\mathrm{T}$. Although the model describes the measured $R_{\mathrm{pPb}}$ at $p_\mathrm{T}<3$\,GeV/$c$, a deviation can be seen at intermediate and high $p_\mathrm{T}$.
\section{Summary}
In this contribution the new ALICE measurements of the $\Lambda^+_{\rm c}$ cross section, $\Lambda^+_{\rm c}/{\rm D^0}$ and $\Lambda^+_{\rm c}$ $R_{\rm pPb}$ down to $p_{\rm T}$ = 0 in p--Pb collisions were shown. The baryon-to-meson ratio is enhanced with respect to $\mathrm{e}^+\mathrm{e}^-$ collisions, indicating that the charm fragmentation is not a universal process across different collision systems. In the future the measurement precision of heavy-flavour hadrons will improve in Run 3 and beyond due to increased data taking rates and upgraded tracking detectors. An improved tracking resolution of the new Inner Tracking System will also allow a better separation of the reconstructed signal from the large combinatorial background \cite{ITSRun3}.

\end{document}